\documentclass[useAMS,usenatbib,usegraphicx]{mn2e}

\title[Dust scattering and the DIG]
{
On the nature of diffuse ionized gas in galaxies -- I\\ The contribution of dust scattering to diffuse line emission
}

\author[Ascasibar et al.]
{
Y.~Ascasibar$^{\star 1,2}$, G.~Guidi$^{3}$, J.~Casado$^{1,2}$, C.~Scannapieco$^{3}$, and A.~I.~D\'{i}az$^{1,2}$
\\
$^{1}$ Departamento de F\'{i}sica Te\'{o}rica, Universidad Aut\'{o}noma de Madrid, Madrid 28049, Spain\\
$^{2}$ Astro-UAM, UAM, Unidad Asociada CSIC\\
$^{3}$ Leibniz Institut f\"{u}r Astrophysik Potsdam, An der Sternwarte 16, Potsdam 14482, Germany
}

\date{\bf Draft version 1.0 (\today)}

\newcommand{\msun}{\ensuremath{{\rm M}_\odot}}

\newcommand{\eqref}[1]{~(\ref{#1})}

\newcommand{\Ha}{H$\alpha$}
\newcommand{\HIIR}{H{\sc ii} regions}
\newcommand{\threshold}{\ensuremath{I_{\rm thr}}}
\newcommand{\rms}{\ensuremath{I_{\rm rms}}}

\usepackage{color}

\begin{document}

\maketitle

\begin{abstract}
In this work, we investigate the contribution of dust scattering to the diffuse \Ha\ emission observed in nearby galaxies.
As initial conditions for the spatial distribution of \HIIR, gas, and dust, we take three Milky Way-like galaxies from state-of-the-art cosmological hydrodynamical simulations that implement different prescriptions for star formation, feedback, and chemical enrichment.
Radiative transfer has been solved \emph{a posteriori}, using the publicly-available Monte Carlo code {\sc sunrise} to take into account dust absorption and scattering of the \Ha\ photons, originating exclusively from the \HIIR.
No contribution from recombinations in the diffuse ionized gas (DIG) component is explicitly or implicitly included in our model.
Our main result is that the flux arising from scattered light is of the order of $1-2$ per cent of the \Ha\ flux coming directly from the \HIIR.
Building upon previous studies, we conclude that the DIG contributes lass than 50 per cent of the total \Ha\ emission.
\end{abstract}


\footnotetext[1]{E-mail: yago.ascasibar@uam.es}

\section{Introduction}

The presence of a warm, low-density, ionized component of the interstellar medium (ISM) has long been detected in the Milky Way through its free-free absorption of background sources \citep{Hoyle&Ellis63}, pulsar dispersion measures \citep{Hewish+68}, and optical emission lines \citep{Reynolds+73}.
Comparison of these measurements along common lines of sight indicate that such diffuse ionized gas (DIG) accounts for a large fraction of the ionized hydrogen in the Galaxy, filling about 30 per cent of the volume within a $\sim 2$~kpc layer around the Galactic mid-plane.

Observations of the \Ha\ recombination line have also revealed the presence of a similar ionized component in most, if not all, nearby galaxies.
There is general agreement that the DIG occupies of the order of 80\% of the projected area of the disk, and that it has a larger scale height than the distribution of classical \HIIR.
Both components contribute roughly equally to the total \Ha\ emission, with lower limits to the DIG contribution varying from 25 to 50 per cent and upper limits around $50-70\%$.
The separation is not straightforward, though, and the results depend somewhat on the adopted definition \citep[e.g.][]{Deharveng+88, Hoopes+96, Shopbell&Bland-Hawthorn98, Zurita+00, Thilker+02, Oey+07}.
Spatial resolution, as well as the presence of low surface brightness structures and/or filamentary \HIIR\ \citep{Dopita+06}, may also play an important role.

The detailed properties of the DIG, as well as its physical origin, are still debated \citep[see e.g.][and references therein]{Haffner+09}.
It is generally believed, based on photon counting arguments, that the diffuse gas is also ionized by massive stars \citep{Reynolds84}, but the energy balance and ionization structure are poorly understood.
By studying the relative intensity of forbidden lines to \Ha, it has been found that, both in the Milky Way as well as in other galaxies, the temperature and ionization conditions of the DIG are different from those typical of \HIIR.
Temperatures are higher in the DIG, and some elements, like Sulphur, display a higher ionized fraction, while others, like Helium, are underionized.
The temperature increases gradually with height above the mid-plane, and there are systematic variations of the line ratios suggesting that the spectrum of the ionizing radiation responsible for the DIG component is modified as it propagates through the ISM.
There is significant spatial correlation between \HIIR\ and diffuse emission \citep[e.g.][]{Ferguson+96, Hoopes+96}, and the line ratios change relatively smoothly from values similar to those observed in \HIIR\ towards much higher intensities of the forbidden lines with increasing distance to the mid-plane \citep[e.g.][]{Reynolds88, Dettmar&Schulz92, Golla+96, Rand98, Haffner+99, Otte+02}.

Although most of the diffuse emission would indeed trace the presence of the DIG component, some fraction arises from light emitted somewhere else (e.g. bright \HIIR\ close to the mid-plane) and is scattered towards the observer by interstellar dust particles regardless of the actual ionization state of the gas.
The fraction of scattered light can be relatively high in the direction of dense clouds at high latitudes, and it may contribute, overall, of the order of $10-40$ per cent of the total diffuse emission \citep[see e.g.][]{Reynolds+73, Jura79, Wood&Reynolds99, Mattila+07, Witt+10, Brandt&Draine12, Ienaka+13, Planck_XXV}.
Moreover, radiative transfer simulations suggest that the observed gradients in line ratios can be understood in terms of dilution of the intrinsic DIG ratios by the scattering of light originating in \HIIR\ \citep[e.g.][]{Ferrara+96, Wood&Reynolds99, Seon&Witt12}, although recent work by \citet{Barnes+15} shows that the fraction of \Ha\ emission arising from scattered light drops below $5-10$ per cent beyond 300~pc from the mid-plane, and that it accounts for less than $10-20$ per cent of the total \Ha\ luminosity of the DIG, depending on the assumed density structure.

Here we consider the contribution of dust scattering to the diffuse line emission observed in nearby galaxies in terms of the luminosity of classical \HIIR.
More precisely, we estimate the fraction the \Ha\ flux that is scattered in the direction of the observer.
Our results are based on radiative transfer calculations that use cosmological numerical simulations to provide realistic conditions for the distribution of gas, dust, and \HIIR.
The production of simulated \Ha\ images is described in Section~\ref{sec_simulations}, the automatic separation of \HIIR\ from diffuse emission as well as their relative contribution to the total \Ha\ flux are discussed in Section~\ref{sec_results}, and our main conclusions are summarized in Section~\ref{sec_conclusions}, where the implications for the physical origin of the DIG are briefly touched upon.

\begin{figure*}
\includegraphics[height=.2\textheight]{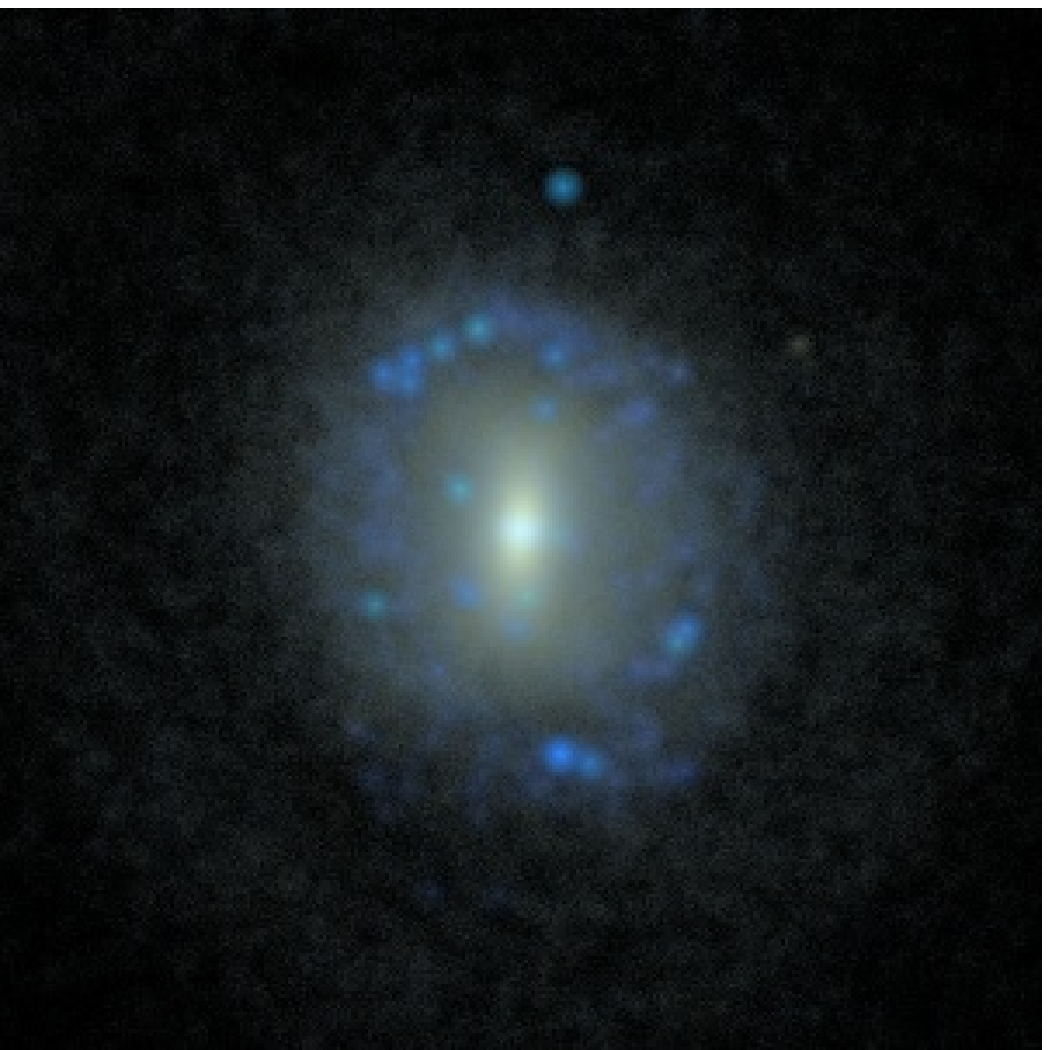} \hfill
\includegraphics[height=.2\textheight]{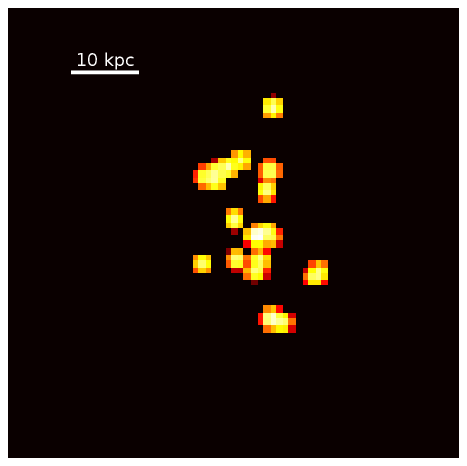} \hfill
\includegraphics[height=.2\textheight]{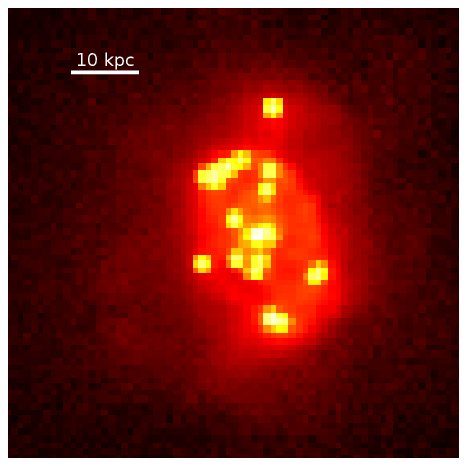} \\
\includegraphics[height=.2\textheight]{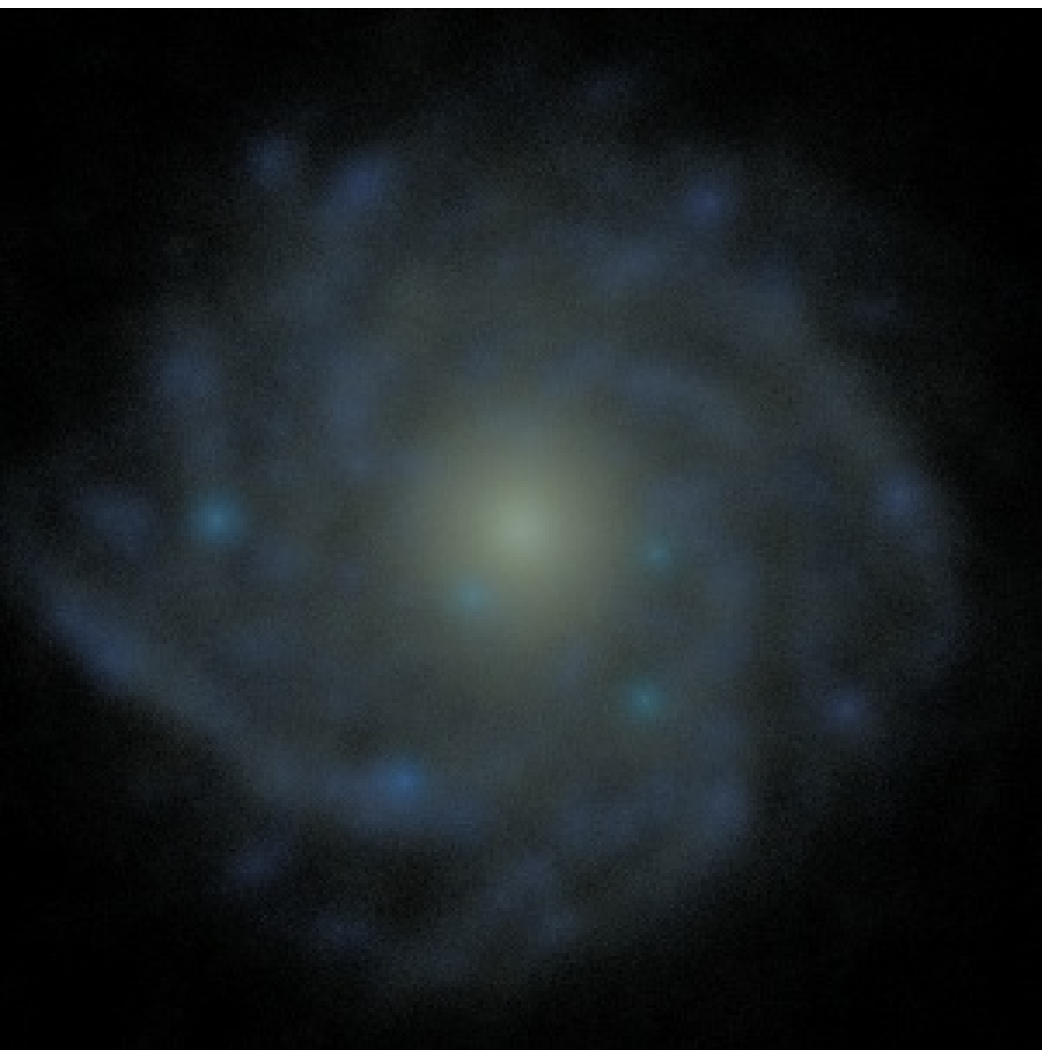} \hfill
\includegraphics[height=.2\textheight]{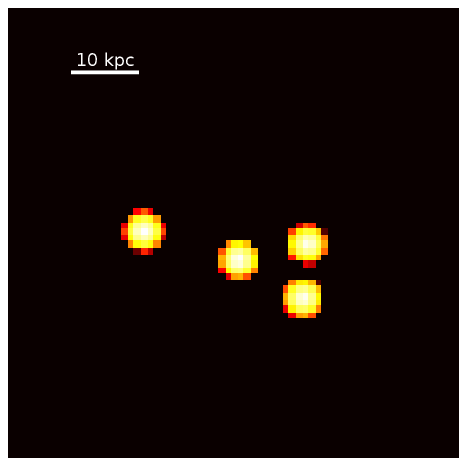} \hfill
\includegraphics[height=.2\textheight]{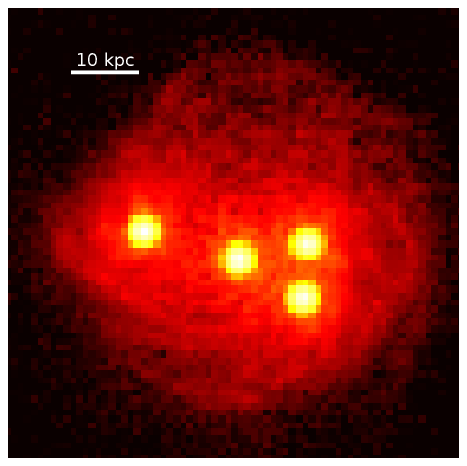} \\
\includegraphics[height=.2\textheight]{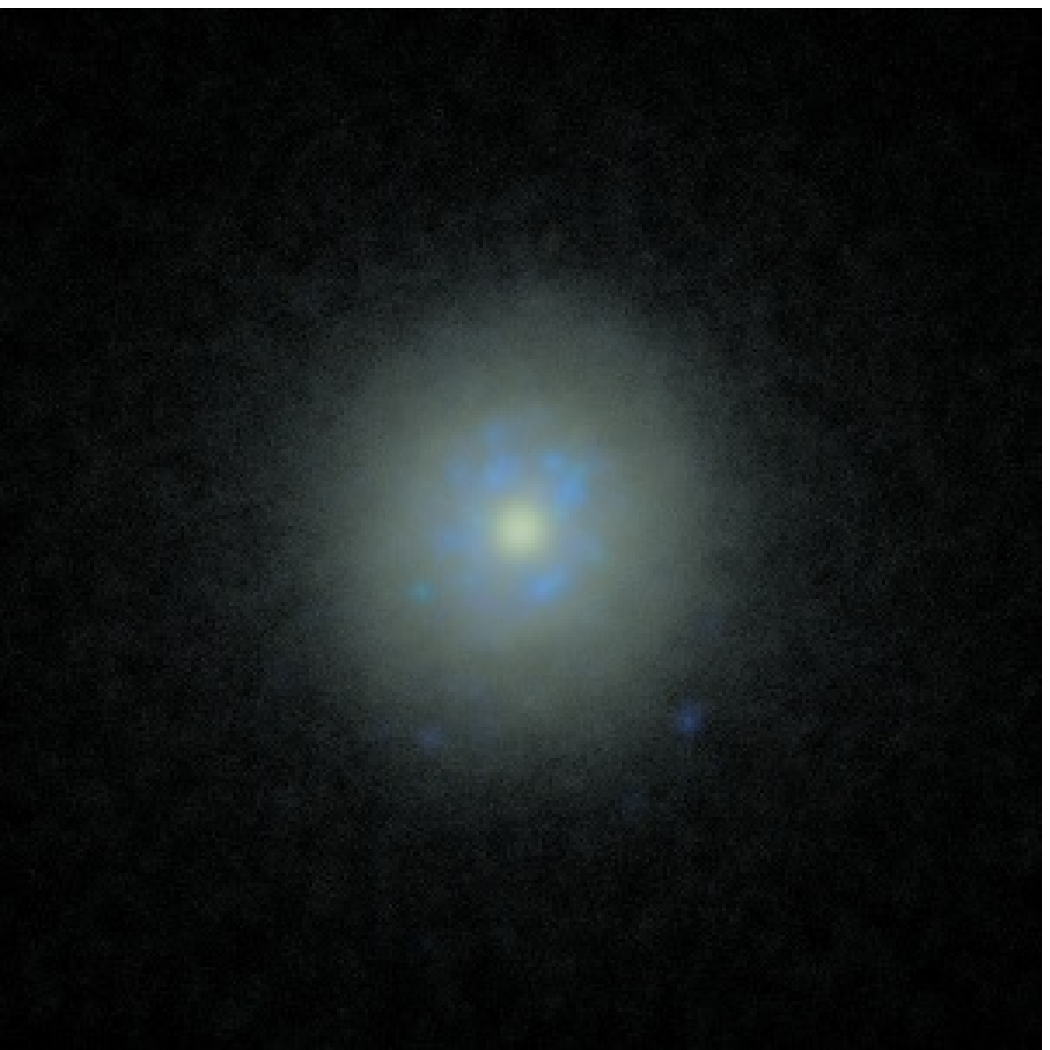} \hfill
\includegraphics[height=.2\textheight]{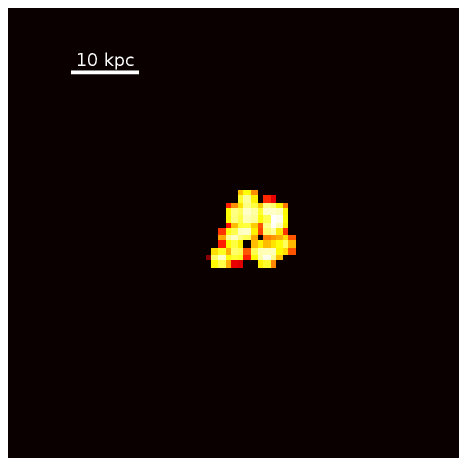} \hfill
\includegraphics[height=.2\textheight]{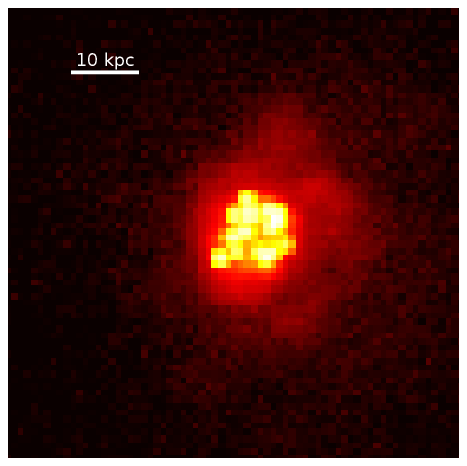} \\
\caption
{
Synthetic images of our three simulated galaxies (from top to bottom, Aquarius A, B, and C) for a $\sim 70 \times 70$-kpc face-on view.
A colour composite based on the broadband SDSS $(u,r,z)$ bands is shown on the left column.
The middle column displays the modelled intensity of the \Ha\ line emitted by the \HIIR\ associated to young stellar particles, ignoring the effects of dust absorption and scattering.
The results obtained by including them are shown on the right column.
}
\label{fig_face}
\end{figure*}
\begin{figure*}
\includegraphics[height=.2\textheight]{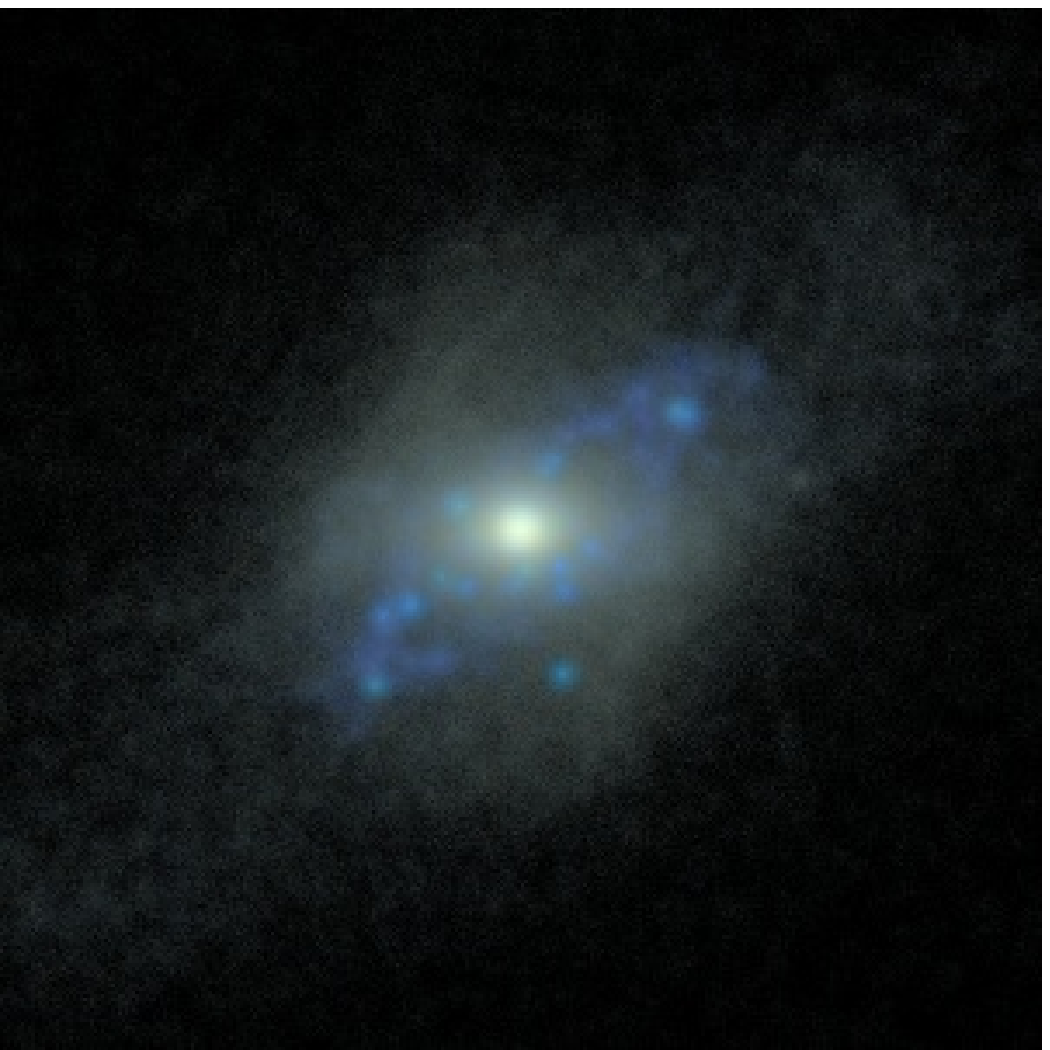} \hfill
\includegraphics[height=.2\textheight]{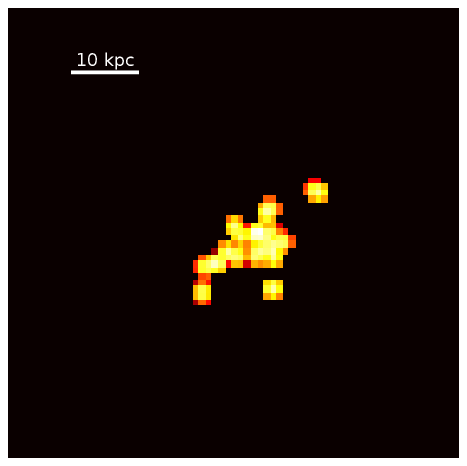} \hfill
\includegraphics[height=.2\textheight]{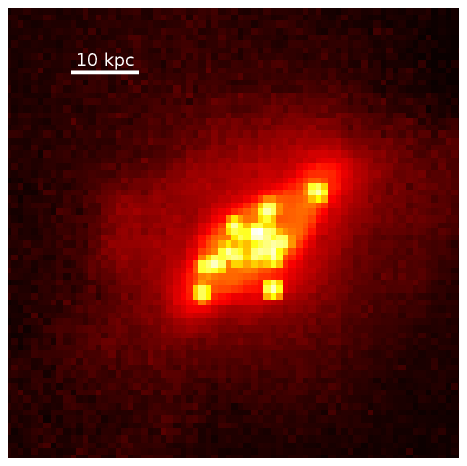}\\
\includegraphics[height=.2\textheight]{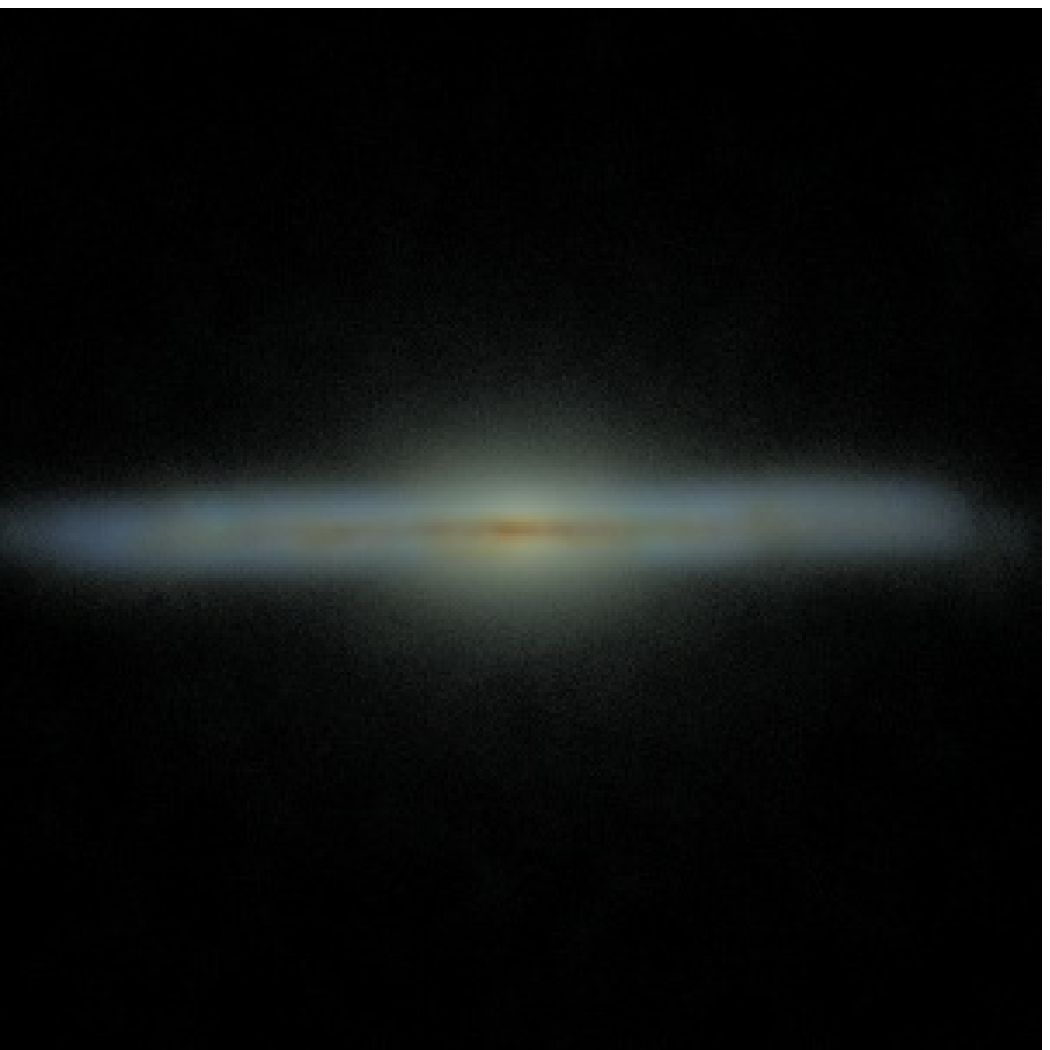} \hfill
\includegraphics[height=.2\textheight]{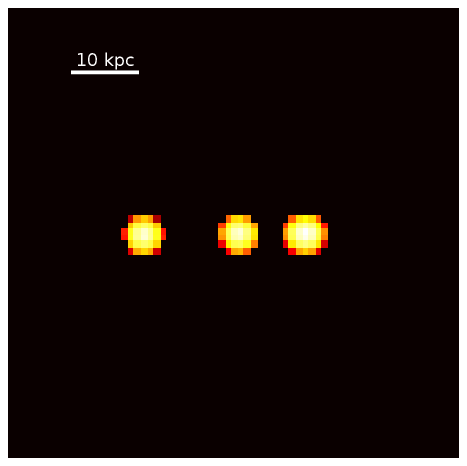} \hfill
\includegraphics[height=.2\textheight]{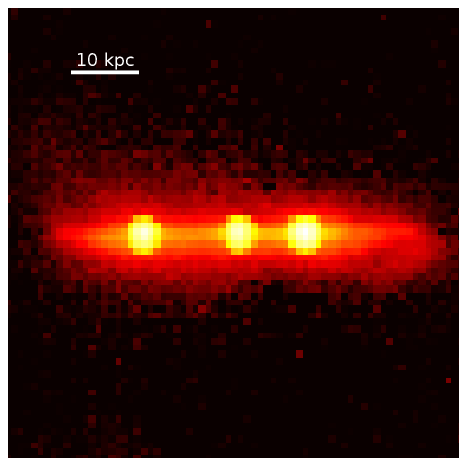}\\
\includegraphics[height=.2\textheight]{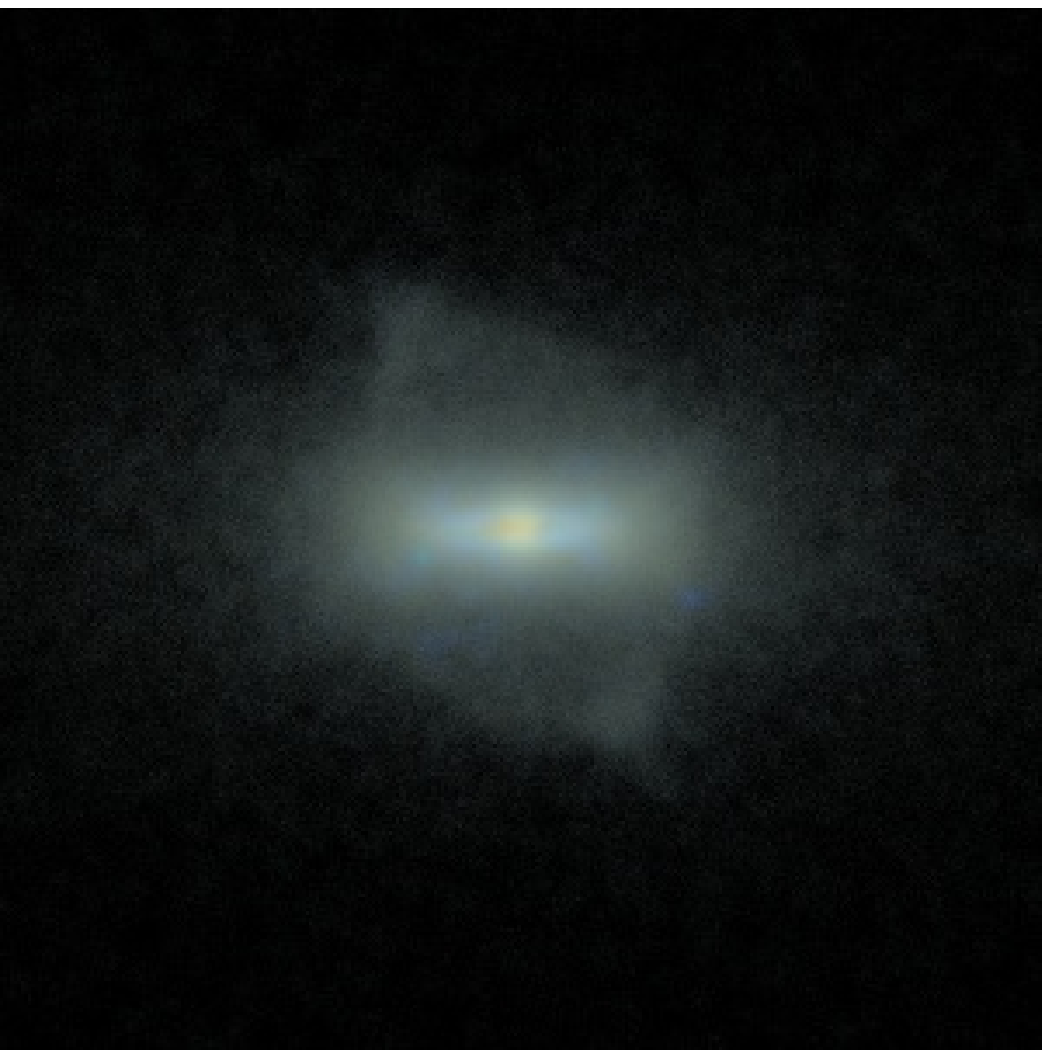} \hfill
\includegraphics[height=.2\textheight]{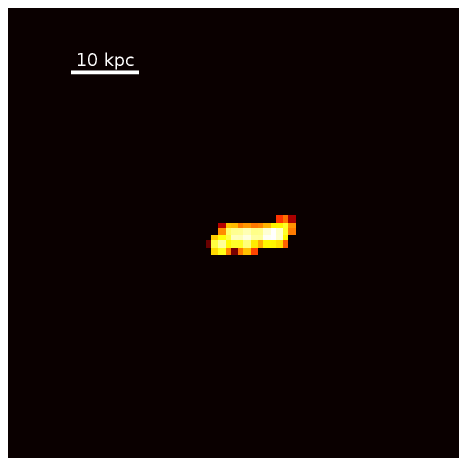} \hfill
\includegraphics[height=.2\textheight]{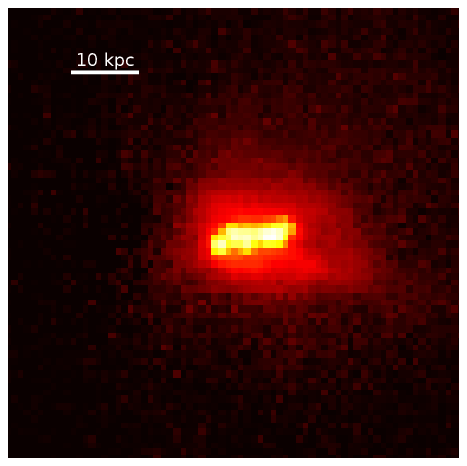}\\
\caption
{
Synthetic images as in Figure~\ref{fig_face}, for an edge-on view.
}
\label{fig_edge}
\end{figure*}

\section{Simulated maps}
\label{sec_simulations}

We use three different hydrodynamical simulations of galaxies in a $\Lambda$CDM universe as models to generate mock H$\alpha$ maps.
The initial conditions for these hydrodynamical simulations are based on the dark-matter-only Aquarius simulation \citep{Springel08}, from which target halos are identified at redshift zero and then resimulated, including
hydrodynamics and stellar physics, using the zoom-in technique \citep{Tormen97}.
The cosmological parameters assumed are
$\Omega_{\rm m} = 0.25 $, $\Omega_{\Lambda} = 0.75$,  $\Omega_{\rm b} = 0.04$, 
$\sigma_8 = 0.9$ and $H_0 = 100 \, h$~km~s$^{-1}$~Mpc$^{-1}$ with $h=0.73$.
All the simulations used in the present work have mass resolution of 
$2-5\times 10^{5}$ M$_\odot$ 
for stellar/gas particles and $1-2\times 10^{6}$ M$_\odot$ for dark matter 
particles, and similar gravitational softenings ($300-700$ pc).

Our three halos were selected as possible candidates for the formation of Milky Way-like galaxies, i.e. they have similar virial mass ($M_{\rm halo} \sim 1-1.5 \times 10^{12} \msun$) and formed in an isolated environment (no neighbour more massive than half of their mass within 1.4~Mpc at redshift zero, see 
\citealt{Scannapieco09}).
Two of them (halos A and C) have been simulated with a new 
version of the 
\citet{Scannapieco05, Scannapieco06} code, which is an implementation of
Supernovae (SNe) feedback and chemical enrichment in the Tree-PM SPH code 
Gadget-3 \citep{Springel05}. The updated version (Poulhazan et al., 
in prep.) keeps the original \citet{Scannapieco05, Scannapieco06} model 
of star formation, thermal feedback from SNe Type Ia and Type II, 
metal-dependent cooling, but uses 
a different Initial Mass Function (Chabrier), updated chemical yields (from 
\citealt{Portinari98}), and includes chemical enrichment from AGB stars.

The third halo (Aquarius B) was simulated with the \cite{Aumer13} update to
the \citet{Scannapieco05, Scannapieco06} model.
The main changes in the chemical model are new yields, the inclusion of metal 
diffusion in the ISM, and the use of a different IMF (Kroupa), as well 
as different metal enrichment from stars in the AGB phase and metal-dependent cooling function.
The energy feedback model in Aumer et al. is different compared to 
Scannapieco et al., as the energy released by SNe
explosions is divided into a thermal and a kinetic part; in addition to SNe, 
the code includes feedback from radiation coming from massive young stars; 
the result of the Aumer et al. update is in general stronger feedback compared
to the original model, and hence younger, more disk dominated and metal rich 
galaxies, although with flatter metallicity gradients \citep{Guidi15}.

The Spectral Energy Distribution (SED) of the simulated galaxies is calculated from the snapshots at
redshift zero using the Monte Carlo radiative transfer code {\sc sunrise} \citep{Jonsson06, Jonsson09} to propagate the light coming from stars, nebulae, and dust through the ISM.

The first step of the radiative transfer simulation is to assign each star 
particle a spectrum according to its age, metallicity and mass.
A nebular spectrum generated with the photo-ionization code 
{\sc mappings III} \citep{Groves04,Groves08} is given to young star particles 
(age $<$ 10 Myr), including 
(sub-grid) the effect of dust absorption and emission within the nebula. 
Most of the {\sc mappings III} parameters are constrained by the 
hydrodynamical simulation,
except the time-averaged fraction of stellar 
cluster solid angle covered by the Photo-Dissociation Region, 
that we fix to $f_{\rm PDR} = 0.2$ 
\citep{Jonsson09}.
Star particles older than 10 Myr are assigned spectra generated with 
STARBURST99 stellar population synthesis model \citep{Leitherer99}, assuming a \citet{Kroupa02} IMF and 
Padova 1994 stellar tracks.

After the spectrum of each stellar particle is calculated, the 
light is 
propagated through the ISM using a Monte Carlo approach \citep{Jonsson06}, 
assuming a constant dust-to-metals ratio of 0.4 \citep{Dwek98}. 
Dust extinction in {\sc sunrise} is implemented by a Milky Way-like 
extinction curve normalized to $R_V =3.1$ \citep{Cardelli89,Draine03}.
To calculate the amount of light scattered by dust grains, 
{\sc sunrise} assumes the \citet{Henyey41} phase function
\begin{equation}
 p(\theta) = \frac{1}{4\pi} \frac{1-g^2}{[1+g^2 - 2g \cos(\theta)]^{3/2}}
\end{equation}
where  $-1 < g < 1$ parameterizes the asymmetry of the phase 
function (see \citealt{Jonsson06} for details).
After the radiative transfer in the dusty ISM, cameras with different
field of view collect the light coming from the galaxy, 
giving the SED in each pixel of the camera (spaxel).
In this calculation we put two cameras for each galaxy, one 
observing face-on and the other edge-on with respect to the star's total angular momentum; in this way we consider the two most relevant cases in terms of dust optical depth.
The spatial resolution of the spaxels is 1 kpc.

To extract the H$\alpha$ maps from {\sc sunrise} datacubes, we first
run ten different random realizations of the same galaxy with {\sc sunrise}, 
in order to reduce the Monte Carlo noise and obtain `noiseless' 
datacubes, averaging the SED in each spaxel over the ten random 
realizations. With the ten random realizations we reach a 
Signal-to-Noise of $\sim 300-400$ in 
the central spaxels and S/N$\sim 5 - 10$ in the outer ones. From the 
noiseless {\sc sunrise} datacubes we extract the H$\alpha$ emission
in each spaxel, after subtracting the stellar continuum calculated 
with {\sc sunrise} not including nebular emission when generating the SED.

Synthetic images of our three galaxies for the face-on and an edge-on views are shown in Figures~\ref{fig_face} and~\ref{fig_edge}, respectively.
In both cases, the left column displays a composite broadband image of each galaxy in the SDSS $(u,r,z)$ bands.
Our three objects are fairly realistic late-type galaxies with different total masses and luminosities, bulge-to-disk ratios, and present-day star formation rates \citep[halos A-CS+, B-MA, and C-CS+ in][]{Guidi15, Guidi+16}.
Star-forming regions can be clearly seen as blue spots associated to the presence of young O and B stars, superimposed over a much smoother (and redder) continuum, dominated by the older stellar population.
Most of these regions are located in the galactic disks, at a certain distance from the galactic centre.
The effect of dust extinction is most evident in the edge-on view of Aquarius B (middle row in Figure~\ref{fig_edge}), where a dust lane obscures the emission near the mid-plane of the disk.

The \Ha\ emission from the young star particles that represent classical \HIIR\ in our simulations is plotted on the middle column of Figures~\ref{fig_face} and~\ref{fig_edge}.
Dust absorption and scattering have been turned off in these images, and therefore they merely trace the \Ha\ emission from the {\sc mappings III} models.
There are only a handful of \HIIR\ in each galaxy, and the distribution of these maps is significantly more clumpy than real galaxies, which usually contain a large number of smaller \HIIR, as well as intrinsic diffuse emission from the DIG.
The number of young stellar particles depends, of course, on the numerical resolution of the simulation, but this does not constitute a critical caveat for our present purposes, since the total \Ha\ flux should be proportional to the integrated star formation rate of the galaxy.
Our simulations do \emph{not} consider any other kind of nebular emission; in particular, there is no contribution from the DIG.

Maps of the \Ha\ emission, taking dust absorption and scattering into account, are displayed on the right column of the figures.
A significant component of extended emission is clearly visible, which is entirely due to the scattering of \Ha\ photons (arising from the \HIIR) by the dust particles in the diffuse ISM.
The scattered light extends throughout the whole disk, and it reaches a considerable height over the galactic plane.
The intensity, however, is more than an order of magnitude lower than the typical values within the \HIIR, and therefore the contribution to the total observed flux from the galaxy is only of the order of a few percent.

\section{Results}
\label{sec_results}

Let us now try to constrain more quantitatively the precise ratio between the \Ha\ emission arising from scattered light and the flux escaping directly from the classical \HIIR.
In order to do so, the first step is to separate diffuse emission (which, in our model, is entirely due to dust scattering) from the bright, compact \HIIR\ that dominate the galaxy luminosity in \Ha.
Here we have a significant advantage with respect to real galaxies, since, to some extent, we already know the solution (the contribution of the \HIIR\ can be clearly seen on the middle columns of Figures~\ref{fig_face} and~\ref{fig_edge}), but even in such a favourable case, finding an objective prescription (that can be automatically implemented as a computer algorithm) to carry out the separation is not trivial in practice.

We have adopted a very simple scheme, choosing an intensity threshold \threshold\ to define the boundary between both components.
Pixels above the threshold will be classified as \HIIR, whereas pixels displaying lower intensities will be assigned to the diffuse component.
The choice of \threshold\ is not straightforward, and it may potentially affect the results, but the impact of the adopted value is rather limited in our simulations because there is a very high contrast between the \Ha\ emission of the \HIIR\ and the diffuse scattered light.
In real galaxies, diffuse \Ha\ emission from hydrogen recombination in the DIG, as well as observational issues (such as e.g. the subtraction of the stellar continuum, or the effect of the point-spread function), would complicate the task, and a more careful analysis would probably be required.

\begin{figure}
\includegraphics[width=.45\textwidth]{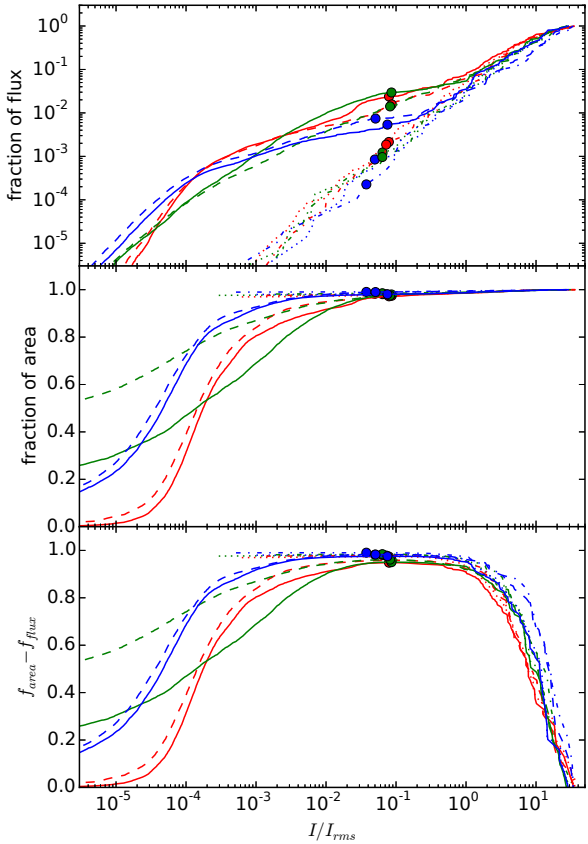}
\caption
{
Relation between the (normalized) intensity threshold defining diffuse emission and its fractional contribution to the total flux (top) and area (center).
The bottom panel shows the difference between both fractions.
We choose \threshold\ at the maximum of this curve (i.e. most of the observed flux coming from a small fraction of the area).
Solid and dashed lines correspond to the face-on and edge-on views, respectively.
Results obtained by neglecting dust scattering are plotted as dotted lines.
Red, green, and blue colours indicate our three different galaxies A, B, and C, respectively.
Coloured circles in all panels mark the values of each curve at \threshold.
}
\label{fig_cumulative}
\end{figure}

Our prescription to choose \threshold\ is illustrated in Figure~\ref{fig_cumulative}.
In order to remove the dependence on the overall normalization of the image, we will express our intensity threshold in terms of the root-mean-square intensity $\rms \equiv \sqrt{\langle I^2 \rangle}$ over the whole field of view.
For images as clumpy as our simulated \Ha\ maps, \rms\ is representative of the typical values of the surface brightness measured within the \HIIR, and thus we would like to argue that it provides a natural scale upon which to base the identification of bright knots against a fainter diffuse background, not only in this context, but maybe also in the general case.

The fraction of flux arising from regions below a given intensity level is plotted on the top panel of Figure~\ref{fig_cumulative}.
In the images without dust scattering (central columns in Figures~\ref{fig_face} and~\ref{fig_edge}, dotted lines in Figure~\ref{fig_cumulative}), the cumulative fraction of flux rises as a power law of the intensity, whereas in the images where diffuse emission from dust scattering is included, one can readily observe a plateau at a level of about 1 per cent for both the face-on (solid lines) and the edge-on (dashed lines) views.

A very different behaviour is observed in the fraction of the area covered by regions below a given intensity level, shown on the middle panel of Figure~\ref{fig_cumulative}.
For most pixels, the intensity of the images without dust scattering is zero, and therefore the dotted curves are practically horizontal lines.
When radiative transfer is included, the cumulative fraction of pixels increases more steadily due to the contribution of the diffuse emission.
Although the qualitative behaviour is the same in all cases, here we find significantly more variety in the shapes of the different curves (different galaxies and orientations) than in the fraction of total flux.
We also expect that a different field of view, zooming into the central region, would also have a sizeable impact on the detailed form of the cumulative fraction of area at the faint end.

Our prescription identifies diffuse emission with a large portion (area) of the image that contributes a rather small fraction of the flux, considering its size (i.e. low surface brightness level).
Conversely, \HIIR\ correspond to bright, compact knots of emission that make a significant (even dominant) contribution to the total flux while covering a very small area of the image.
Therefore, we plot on the bottom panel of Figure~\ref{fig_cumulative} the difference between the cumulative fraction of area $f_{\rm area}$ with intensity smaller than $I$ and the corresponding fraction of flux $f_{\rm flux}$.
For a completely uniform image, with all points having exactly the same intensity, both fractions would be step functions, and the difference $f_{\rm area}-f_{\rm flux}$ would vanish for any value of the intensity.
If the image was composed of two discrete levels, with intensities $I_0<I_1$, occupying areas $A_0>A_1$, respectively, $f_{\rm area}-f_{\rm flux}$ would display a plateau in the interval between $I_0$ and $I_1$, being identically zero otherwise.
The value attained in the plateau would in fact provide a quantitative measure of the clumpiness of the image, ranging from zero for a uniform image to almost unity when all the emission is concentrated in a single pixel.

\begin{table}
\begin{center}
\begin{tabular}{cccccc}
\hline
Galaxy & $F_0$ & Orientation & $F_{\rm obs}$ & $\frac{ I_{\rm thr} }{ \rms }$ & $\frac{ F_{\rm sca} }{ F_{\rm HII} }$ \\
\hline
A & 2.063 & face-on & 2.036 & 0.079 & 0.024 \\
  &       & edge-on & 1.858 & 0.087 & 0.016 \\
B & 0.226 & face-on & 0.231 & 0.086 & 0.030 \\
  &       & edge-on & 0.144 & 0.082 & 0.014 \\
C & 1.451 & face-on & 1.454 & 0.075 & 0.005 \\
  &       & edge-on & 0.970 & 0.051 & 0.007 \\
\hline
\end{tabular}
\end{center}
\caption
{
First two columns identify the galaxy and provide the total flux $F_0$ of the \Ha\ line emitted by its \HIIR.
Then, for each orientation, we quote the final observed flux $F_{\rm obs}$ integrated over the whole galaxy, including dust absorption and scattering, the intensity threshold \threshold\ (in units of the root-mean-square value) adopted to separate compact \HIIR\ from diffuse emission, and ratio between the fluxes arising from each component (recall that, in our model, diffuse emission is entirely due to dust scattering).
}
\label{tab_flux}
\end{table}

This is exactly the trend observed in our simulated images.
Of course, the intensity is not perfectly uniform neither within the \HIIR\ nor in the area illuminated by dust scattering, and therefore the plateau is not strictly flat, and its edges are not vertical lines.
However, as can be seen on the top panel of the figure, there is relatively little difference in the fraction of flux contributed by each component if one chooses any threshold intensity within the interval where $f_{\rm area}-f_{\rm flux}$ is approximately constant.
We have selected our value of \threshold\ exactly at the maximum, which happens slightly below ten per cent of the root-mean-square intensity of the image for all galaxies and orientations, regardless of whether dust scattering has been included or not.
The location of our adopted division between the \HIIR\ and the extended diffuse emission is indicated by the coloured circles in Figure~\ref{fig_cumulative}, and precise numerical values are quoted in Table~\ref{tab_flux}.
We find that, for our simulated galaxies, our automatic procedure yields $\threshold\sim 0.08\,\rms$.

We have verified by visual inspection that this simple scheme is able to reliably identify the \HIIR\ in the simulated images.
In particular, only a few pixels, representing about one thousandth of the total flux (see top panel of Figure~\ref{fig_cumulative}), are assigned to the diffuse component in the images without dust scattering.
On the other hand, practically the same regions are recovered in the images with radiative transfer, suggesting that a refined version of our algorithm could be applied to separate \HIIR\ and diffuse emission in observational data.

According to our results (summarized in the last column of Table~\ref{tab_flux}), diffuse emission (i.e. dust scattering) contributes of the order of $1-2$ per cent of the \Ha\ line emission arising from compact \HIIR.
The exact fraction, of course, depends on the morphology, metallicity, and orientation of the galaxy, but for the objects we considered it does not seem to vary by more than a factor of two or three.

From the numbers in Table~\ref{tab_flux}, we also conclude that, when an object is observed face-on, dust attenuation is very small, also of the order of a few per cent, and thus the total flux of the \Ha\ line may even increase with respect to the intrinsic luminosity of the \HIIR\ due to the contribution of dust scattering.
For the edge-on views, the light from the \HIIR\ is more heavily absorbed, and the net flux that reaches the observer may decrease by as much as 30 per cent.
Although our statistics is certainly very limited, we do not observe any systematic trend in the ratio between the scattered light and the compact emission as a function of orientation; in galaxies A and B, the diffuse component is more prominent in the face-on view, whereas the opposite is true for our simulated images of Aquarius-C.

\section{Discussion and conclusions}
\label{sec_conclusions}

In this work, we study the contribution of dust scattering to the \Ha\ luminosity of a Milky Way-like galaxy in the local universe.
More precisely, we solve the radiative transfer of \Ha\ light, emitted by classical \HIIR, through the diffuse ISM of three galaxies extracted from cosmological hydrodynamical simulations.

Based on the ratios quoted on the last column in Table~\ref{tab_flux}, we conclude that the flux arising from dust scattering $F_{\rm sca}$ is of the order of $1-2$ per cent of the emission $F_{\rm HII}$ from compact \HIIR.
Although we have not included the DIG in our modelling, it is possible to obtain meaningful constraints on its physical origin by combining our main result
\begin{equation}
 \frac{ F_{\rm sca} }{ F_{\rm HII} } \sim 0.015 \pm 0.01
\end{equation}
with previous findings in the literature.

In principle, one may expect that most of the ionizing flux from young O and B stars is absorbed by the surrounding gas, creating the classical \HIIR, and that only a small fraction (of the order of a few per cent) escapes from the galaxies \citep[see e.g.][]{Barger+13, Kim+13}.
Assuming that the DIG is produced by the absorption of leaking UV radiation, the required fraction of photons $f_{\rm DIG}$ could vary between 20 and 80 per cent depending on the luminosity of the \HIIR\ \citep[e.g.][]{Zurita+02}.
Under this assumption,
\begin{equation}
 \frac{ F_{\rm DIG} }{ F_{\rm HII} } \sim \frac{ f_{\rm DIG} }{ 1-f_{\rm DIG} }
\end{equation}
and one trivially obtains
\begin{equation}
 \frac{ F_{\rm HII} }{ F_{\rm obs} } = \frac{ F_{\rm HII} }{ F_{\rm HII} + F_{\rm sca} + F_{\rm DIG} }
 \sim \frac{ 1 }{ 1.015 + \frac{ f_{\rm DIG} }{ 1-f_{\rm DIG} } }
\end{equation}
as well as
\begin{equation}
 \frac{ F_{\rm sca} }{ F_{\rm DIG} }
 \sim 0.015\ \frac{ 1-f_{\rm DIG} }{ f_{\rm DIG} }
\end{equation}

If $f_{\rm DIG} \sim 0.5$, the emission from the compact and the diffuse ionized components would be of the same order, and the contribution of dust scattering, according to our model, would be a correction of about $1-2$ per cent with respect to the DIG emission.
This is roughly consistent, but slightly lower than the results reported by \citet{Barnes+15}, where $\frac{ F_{\rm sca} }{ F_{\rm DIG} }$ may be as high as 10 or even 20 per cent in some regions.
If such values were representative of real galaxies, we would advocate leakage fractions of the order of $f_{\rm DIG} \sim 0.1-0.2$, implying that the majority of the \Ha\ emission arises from classical \HIIR.

 \section*{Acknowledgments}

Financial support for the present work has been provided by research grant AYA2013-47742-C4-3-P from the \emph{Ministerio de Econom\'{i}a y Competitividad} (Mineco, Spain) as well as the exchange programmes `Study of Emission-Line Galaxies with Integral-Field Spectroscopy' (SELGIFS, FP7-PEOPLE-2013-IRSES-612701, funded by the Research Executive Agency of the EU) and the Acciones Conjuntas Hispano-Alemanas (PPP-Spain-57050803) promoted by the \emph{Deutscher Akademischer Austausch Dienst} (DAAD, Germany).
YA is supported by contract RyC-2011-09461 of the \emph{Ram\'{o}n y Cajal} programme (Mineco, Spain).
GG and CS acknowledge support from the Leibniz Gemeinschaft, through project SAW-2012-AIP-5 129, and from the High Performance Computer in Bavaria (SuperMUC) through project pr94zo.

 \bibliographystyle{mn2e}
 \bibliography{references}

\end{document}